\documentclass[epsfig]{article}
\usepackage{amsfonts}
\usepackage{graphicx}
\usepackage{amsmath}

\input epsf.sty
\newtheorem{theorem}{Theorem}
\newtheorem{acknowledgement}[theorem]{Acknowledgement}

\makeatletter \@addtoreset{equation}{section}
\renewcommand{\theequation}{\thesection.\arabic{equation}}
\input{tcilatex}
\begin{document}

\title{%
\rightline{\mbox{\normalsize
{Lab/UFR-HEP0502/GNPHE/0502/VACBT/0502}}} \textbf{Hyperbolic Invariance in
Type II Superstrings}}
\author{El Hassan Saidi\thanks{%
h-saidi@fsr.ac.ma} \\
{\small \textit{1 }} {\small \textit{GNPHE, Groupement National Physique des
Hautes Energies, Facult\'{e} des Sciences de Rabat, Morocco.}}\\
{\small \textit{2-Virtual African Centre for Basic Science and Technology,
VACBT\thanks{%
A GNPHE project for the establishement of an African Centre for Basic
Science \& Technology in North Africa.}, Focal point, LabUFR-PHE, Rabat,
Morocco.}}}
\maketitle

\begin{abstract}
We first review aspects of Kac Moody indefinite algebras with particular
focus on their hyperbolic subset. Then we present two field theoretical
systems where these structures appear as symmetries. The first deals with
complete classification of $\mathcal{N}=2$ supersymmetric CFT$_{4}$s and the
second concerns the building of hyperbolic quiver gauge theories embedded in
type IIB superstring compactification of Calabi-Yau threefolds. We show,
amongst others, that $\mathcal{N}=2$ CFT$_{4}$s are classified by Vinberg
theorem and hyperbolic structure is carried by the axion modulus.

\end{abstract}

\tableofcontents

\newpage

\newpage

\section{Introduction}

Kac Moody (KM) algebras $\mathbf{g}$ and their representations have been one
of the basic tools in establishing strong results in quantum field and
superstring theories. These particular bosonic algebras are generally
divided into three basic sectors \cite{km}: (\textbf{1}) Usual finite
dimensional Lie algebras classified by Cartan; they play a crucial role in
Yang-Mills gauge theories and in the understanding of the dynamics of
interacting elementary particles. (\textbf{2}) Infinite dimensional affine
KM algebras classified by Kac and Moody; they play a basic role in
describing the physics of 2d scale invariant systems, in particular string
world sheet dynamics \cite{pol} and a large class of statistical mechanics
of 2d critical phenomena \cite{car,kss}. (\textbf{3}) Indefinite KM algebras
whose role in quantum physics is still unclear although they appear from
time to time in physical literature as possible underlying symmetries in
some specific models \cite{julia}-\cite{Nic3}. Besides of an apparent non
unitary physical behaviour of hypothetical field theoretical models having
indefinite algebras as symmetries, the little interest into these kind of
systems might be also due to lack of complete mathematical results in this
matter. To our understanding, the second reason is the most probable.

The aim of this study is to first make an excursion into indefinite sector
of KM algebras. Then describe two examples where indefinite, in particular
hyperbolic, KM algebras seem to appear as a physical invariance at least
from theoretical point of view. These examples concerns:

(\textbf{a}) The \textit{full} classification of $\mathcal{N}=2$ CFT$_{4}$
using geometric engineering method. As we will see, there are \textit{three}
classes of $\mathcal{N}=2$ CFT$_{4}$, one of them is classified by the
indefinite subset of KM algebras.

(\textbf{b}) Embedding hyperbolic quiver gauge theories in type IIB
superstring compactification on a specific class of K3 fibered Calabi-Yau
threefolds (CY3). We show that the hyperbolic structure is captured by the
axion field of ten dimensional\footnote{%
I thank S Seikh-Jebbari for discussions on the general 4D solutions.} type
IIB string. Axion field carries therefore a trace on some hypothetical
hidden indefinite KM (hyperbolic) symmetries in type IIB string.

The presentation of this paper is as follows: In section 1, I review briefly
classifications of KM algebras. I give two theorems, one on the Vinberg
classification of KM algebras and the other on the W. Li classification of
their hyperbolic subset. In section 3, I expose aspects on geometric
engineering method of 4d supersymmetric quiver gauge theories; in particular
the engineering of fundamental matter. In section 4, I study the
correspondences between the triplet: (\textbf{i}) roots $\alpha _{i}$ of KM
algebras, (\textbf{ii}) 2-cycles $C_{i}$ of ADE geometries of CY3s and (%
\textbf{iii}) gauge coupling moduli $g_{i}$ in quiver gauge theories
embedded in type II strings on CY3s. In section 5, I present two field
theoretic systems where indefinite KM algebras appear as invariances. In
section 6, I give a conclusion.

\section{Classification theorems}

To begin, let us recall some standard tools on Lie algebras, in particular
roots, Cartan matrices and Dynkin diagrams. Given K, a symmetrisable square
matrix with integers entries taken as follows,
\begin{equation}
K_{ii}=2;\qquad K_{ij}<0,\qquad K_{ij}=0\qquad \Rightarrow \qquad K_{ji}=0,
\label{ne}
\end{equation}%
one generally associates a KM algebra $g=g\left( K\right) $ with the triplet
$(\mathbf{\hbar },\Pi ,\Pi ^{\nu })$ realization defining respectively: (%
\textbf{i}) Commuting Cartan subalgebra generators $\left\{ {H_{i}}\right\} $%
, (\textbf{ii}) the basis of simple roots $\left\{ {\alpha }{_{i}}\right\} $
and ( \textbf{iii}) the basis of their coroots $\left\{ \alpha
_{i}^{v}\right\} $. Fo the case where K is symmetrisable, the situation to
be considered here, KM algebras $g$ admit an invariant symmetric bi-linear
form: $\left( ,\right) :g\times g\longrightarrow C$. In terms of this form,
Cartan matrix K is realized as usual as
\begin{equation}
K_{ij}=<\alpha _{i}^{v},\alpha _{j}>=2\frac{(\alpha _{i},\alpha _{j})}{%
(\alpha _{i},\alpha _{i})}.
\end{equation}%
This relation reduces to $\alpha _{i}.\alpha _{j}$ for simply laced algebras
for which there is one root length $\mathbf{(}\alpha ,\alpha \mathbf{)}=2$.
To fix ideas, I will focus my attention below this particular class.

\subsection{Vinberg classification}

Given above Cartan matrix $K$ and so the corresponding KM algebra $g\left(
K\right) $, there is a powerful way to classify these matrices and algebras $%
g\left( K\right) $. This way, first introduced by Vinberg and then developed
by KM, is a very constrained classification method. Its power follows from
the fact that it is based on the existence of two positive vectors $\mathbf{u%
}$ and $\mathbf{v}$ (i.e several positive definite numbers $u_{i},v_{i}>0$)
as required by following theorem.

\textbf{Theorem: \qquad }A generalized indecomposable Cartan matrix $\mathbf{%
K}$ obeys \textit{one and only one} of the following three statements%
\footnote{%
The content of the theorem is formulated in terms of appropriate notations
which are helpful for later use.}: \newline
(\textbf{1})\textbf{\ Finite type Lie algebras: }\textrm{g}$_{finite}$%
\textit{\ ( }$\det \mathbf{K}^{+}>0$ ):

This is the subset of usual finite dimensional Lie algebras \textrm{g}$%
_{finite}$; it is characterized by the existence of two real positive
definite vectors $\mathbf{u=}\left( u_{1},...,u_{r}\right) $ and $\mathbf{v=}%
\left( v_{1},...,v_{r}\right) $ ($u_{i},v_{i}>0$) such that
\begin{equation}
\mathbf{K}_{ij}^{+}u_{j}=v_{j}>0.  \label{a}
\end{equation}%
The upper index on $\mathbf{K}^{+}$ is \ introduced for convenience, its
role will be understood in a moment.\newline
(\textbf{2})\textbf{\ Affine type KM algebras}\textit{, }\textrm{g}$%
_{affine} $ co-rank$\left( \mathbf{K}^{0}\right) =1$, $\det \mathbf{K}^{0}=0$%
\textit{,}\newline
For \textrm{g}$_{affine}$, there exist a unique, up to a multiplicative
factor $\sigma $, \textit{positive integer} definite vector $\mathbf{u}%
=\sigma \mathbf{d}$ ( $u_{i},d_{i}>0$) such that,
\begin{equation}
\mathbf{K}_{ij}^{0}u_{j}=\sigma \mathbf{K}_{ij}^{0}d_{j}=0.  \label{b}
\end{equation}%
This relation means that the affine Cartan matrix $K_{affine}$ has a
vanishing eigenvalue. The $d_{i}$ integers are known as Dynkin weights.
\newline
(\textbf{3}) \textbf{Indefinite type KM algebras, }\textrm{g}$_{indefinite}$%
\newline
For this class of KM algebras, and like for \textrm{g}$_{finite}$, there
exist two real positive definite vectors $\mathbf{u=}\left(
u_{1},...,u_{r}\right) $ and $\mathbf{v=}\left( v_{1},...,v_{r}\right) $
such that,
\begin{equation}
\mathbf{K}_{ij}^{-}u_{j}=-v_{i}<0.  \label{c}
\end{equation}%
As our present study relies on consequences of this basic theorem, let make
four comments which will be used later. (\textbf{a}) Eqs(\ref{a}-\ref{c})
may be combined into a unique relation as follows%
\begin{equation}
\mathbf{K}_{ij}^{q}u_{j}=qv_{i},  \label{d}
\end{equation}%
where $u_{i}$ and $v_{i}$ are as in Vinberg theorem and where $q=+1,0,-1$.
Quantum number $q$ indexes then KM sectors. (\textbf{b}) For indefinite KM
algebras, the sign of the determinant of $\mathbf{K}^{-}$ is indefinite; it
may be either positive, zero or negative. This property is one of the
features reflecting the difficulty in handling indefinite sector of KM
algebras. (\textbf{c}) Indefinite feature of \textrm{g}$_{indefinite}$ is
also encoded by the negative sign of eq(\ref{d}), which carries an
indication on the fact bilinear from $\left( ,\right) $ has an indefinite
signature. Roots $\alpha $ of indefinite KM algebras are in general as
follows,
\begin{equation}
\left( \alpha ,\alpha \right) =2,1;\qquad \qquad \left( \alpha ,\alpha
\right) =2,1,0;\qquad \qquad \left( \alpha ,\alpha \right) \leq 2;
\end{equation}%
(\textbf{d}) Unlike \textrm{g}$_{finite}$ and \textrm{g}$_{affine}$,
classification of indefinite sector is still an open question, except for
some special cases where one disposes of partial results. The subset of
hyperbolic algebras is one of these; it includes: (\textbf{i}) Standard
hyperbolic algebras containing affine KM as a codimension one subalgebra and
(\textbf{ii}) strictly hyperbolic ones containing only finite algebras as
codimension one subalgebras. Examples of such specific subsets of KM
algebras are given by the following $H\widehat{A}_{2}$ and $HB_{2}$\ Dynkin
diagrams,\bigskip

\begin{figure}[tbh]
\begin{center}
\epsfxsize=5cm \epsffile{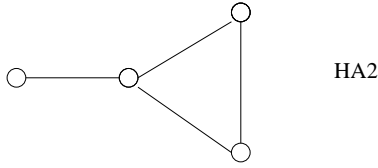}
\end{center}
\caption{\textit{Dynkin diagram of H}$\widehat{\mathit{A}}_{2}$\textit{; the
hyperbolic extension of affine }$\widehat{\mathit{A}}_{2}$\textit{. }}
\label{fig1}
\end{figure}
\begin{figure}[tbh]
\begin{center}
\epsfxsize=3cm \epsffile{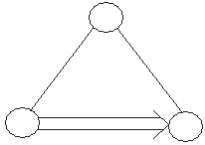}
\end{center}
\caption{\textit{Dynkin diagram of HB}$_{2}$\textit{; the hyperbolic
extension of finite dimensional B}$_{2}$\textit{. }}
\label{fig2}
\end{figure}
Along with these two classes there are other subsets, in particular the so
called extended hyperbolic algebras. Like for the strictly hyperbolic case,
these extensions will not be examined in details in this contribution; so
forget about them and focus on the special subset of hyperbolic algebras
with codimension one affine KM subalgebras. For more information on the
other issues, see \cite{ms1} and refs therein.

\subsection{W. Li classification}

According to W. Li (WL) classification \cite{li}, there are $238$ hyperbolic
Lie algebras containing the following special simply laced list,

\begin{eqnarray}
&&\mathcal{H}_{1}^{4},\quad \mathcal{H}_{2}^{4},\quad \mathcal{H}%
_{3}^{4},\quad \mathcal{H}_{1}^{5},\quad \mathcal{H}_{8}^{5},\quad \mathcal{H%
}_{1}^{6},\quad \mathcal{H}_{5}^{6},\quad \mathcal{H}_{6}^{6},\quad \mathcal{%
H}_{1}^{7},  \notag \\
&&\mathcal{H}_{1}^{7},\quad \mathcal{H}_{1}^{8},\quad \mathcal{H}%
_{4}^{8},\quad \mathcal{H}_{5}^{8},\quad \mathcal{H}_{1}^{9},\quad \mathcal{H%
}_{4}^{9},\quad \mathcal{H}_{5}^{9},\quad \mathcal{H}_{1}^{10},\quad
\mathcal{H}_{4}^{10},
\end{eqnarray}%
where $\mathcal{H}$ refers to hyperbolic, the upper index stands for the
rank of the hyperbolic algebra and the lower one for classification. Let us
give details on some specific examples: The first example we give is $%
\mathcal{H}_{3}^{4}$ to be denoted also as $H\widehat{A}_{2}$. This is the
simplest hyperbolic extension of affine KM algebra $\widehat{A}_{2}$ which
appears as a particular subalgebra. $H\widehat{A}_{2}$ has four simple roots
denoted as $a_{-1}$, $a_{0}$, $a_{1}$ and $a_{2}$ generating all other
roots. The full set $\Delta _{hyp}\left( H\widehat{A}_{2}\right) $ of roots
of $H\widehat{A}_{2}$ contains as a proper subset the roots of affine $%
\widehat{A}_{2}$\ namely,
\begin{align}
& \pm \alpha _{1},\qquad \pm \alpha _{2},\qquad \pm \left( \alpha
_{1}+\alpha _{2}\right) ,  \notag \\
& n\delta ,  \label{ha2} \\
& n\delta \pm \alpha _{1},\qquad n\delta +\alpha _{2},\qquad n\delta \pm
\left( \alpha _{1}+\alpha _{2}\right)  \notag
\end{align}%
where the first line gives the usual roots of ordinary $A_{2}$ and where $%
\delta $ is the familiar imaginary root of affine KM algebras. In eqs(\ref%
{ha2}) $n$ a non zero integer. The $H\widehat{A}_{2}$ algebra is a rank four
KM algebra and has a $4\times 4$ Cartan matrix given by,

\begin{equation}
\mathbf{K}\left( HA_{2}\right) =\left(
\begin{array}{cccc}
2 & -1 & 0 & 0 \\
-1 & 2 & -1 & -1 \\
0 & -1 & 2 & -1 \\
0 & -1 & -1 & 2%
\end{array}%
\right) .
\end{equation}%
Its Dynkin diagram, which is given by figure 1 has four nodes; it is
composed of the two ordinary ones, the affine node and the hyperbolic one.
The second example, we give concerns $H\widehat{D}_{4},$ the hyperbolic
extension of affine $\widehat{D}_{4}$. Its Dynkin diagram has six nodes;
four ordinary ones, one affine and one hyperbolic as shown on figure 3.

\begin{figure}[tbh]
\begin{center}
\epsfxsize=5cm \epsffile{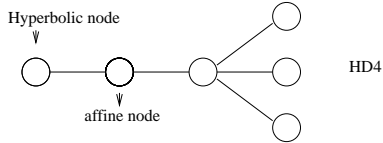}
\end{center}
\caption{\textit{Hyperbolic extension of affine }$\widehat{D}_{4}$.
{\protect\small \textit{{N}ode on left corresponds to hyperbolic extension. }%
}}
\label{fig3}
\end{figure}
Before proceeding, it is interesting to note that some algebras appearing in
the WL classification appears as particular elements in the so called T$%
_{p,q,r}$ algebras. Recall that Dynkin diagrams of $T_{p,q,r}$ involve 3
ordinary A$_{p}$, A$_{q}$ and A$_{r}$ Dynkin chains glued at a trivalent
vertex. Exceptional algebras are the simplest examples; for instance $E_{10}$
indefinite KM algebra is just $E_{10}=T_{7,3,2}$; see also figure 4.\bigskip

\begin{figure}[tbh]
\begin{center}
\epsfxsize=7cm \epsffile{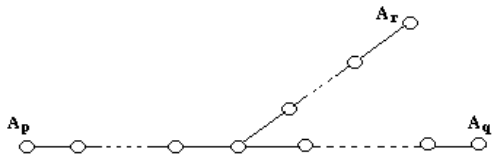}
\end{center}
\caption{{\protect\small \textit{A typical vertex in trivalent algebras T}}$%
_{p,q,r}${\protect\small \textit{\ \ To the central node, it is attached
three legs of A type. }}}
\label{fig4}
\end{figure}
The next note we give concerns useful aspects of T$_{p,q,r}$ algebras; they
will be used later on. On of these aspects is the value of the determinant
Cartan matrix of T$_{p,q,r}$; which can be easily shown to be given by,
\begin{equation}
\det K\left( T_{p,q,r}\right) =pqrC_{3};\qquad C_{3}=\frac{1}{p}+\frac{1}{q}+%
\frac{1}{r}-1.  \label{e}
\end{equation}%
This a remarkable quantity for the study of classification of T$_{p,q,r}$
algebras. Using this relation, one can recover results on the classification
of finite and affine KM algebras. They are respectively associated with
solving the necessary and sufficient conditions $C_{3}>1$ and $C_{3}=0$.
Indefinite $T_{p,q,r}$ algebras corresponds to $C_{3}<1$. It should be noted
also that eq(\ref{e}) is in fact a particular example of a more general
situation. Denoting $T_{p,q,r}$ as $T_{\left[ 3\right] }$ and following \cite%
{sa}, the determinant $\det K\left( T_{\left[ n\right] }\right) $ of
n-dimensional vertex algebra $T_{\left[ n\right] }$ reads as follows:%
\begin{equation}
\det K\left( T_{\left[ n\right] }\right)
=C_{n}\dprod\limits_{i=1}^{n}p_{i};\qquad C_{n}=\left( 2-n\right)
+\sum_{i=1}^{n}\frac{1}{p_{i}},
\end{equation}%
where the $p_{i}$s are non zero positive integers. As one see, $C_{n}$ is
negative for large $n$. \ For a more general formula regarding values of the
determinant of Cartan matrices of indefinite KM algebra and the arithmetic
that governs these computations can be found in \cite{sa}. Note finally that
the conditions%
\begin{equation}
C_{n}\geq n-2;\qquad n=2,3,...,
\end{equation}%
are very restrictive contrary to the constraint eq,
\begin{equation}
C_{n}<n-2;\qquad n=2,3,...,
\end{equation}%
which has infinitely many solutions and so infinitely many series of
indefinite KM algebras.

\section{Geometric engineering of $\mathcal{N}=2$ QFT$_{4}$s}

Geometric engineering of $\mathcal{N}=2$ supersymmetric QFT$_{4}$s is a
tricky method \cite{kmv1,kmv2,bs2,bs3,bs4} to encode superfields contents of
quiver gauge theories embedded in type IIA string compactification on
Calabi-Yau threefolds with ADE geometries. In this picture, gauge and
adjoint matter are associated with nodes of ADE Dynkin diagrams and
bi-fundamental matter with links between nodes.

\textbf{Result}:\qquad In $\mathcal{N}=1$ supersymmetric language, Gauge
multiplets, adjoint matter and bi-fundamental one are naturally encoded by
Dynkin diagrams of finite and affine KM algebras. But what about fundamental
matter transforming in the fundamental representation of $G_{gauge}\times
G_{flavor}$ symmetry ?

\subsection{Adding fundamental matter}

Adding fundamental matter in $\mathcal{N}=2$ QFT$_{4}$s requires more than
Dynkin diagrams; it needs the so called trivalent geometry. This is the
other reason behind our previous interest into $T_{p,q,r}$ KM algebras, the
first one concerned examples of indefinite KM algebras. $T_{p,q,r}$ Dynkin
diagrams have an extra third chain which may be used to encode fundamental
matter. However this is not a soft operation; there is a price one should
pay for engineering fundamental matter in term of trivalent geometry. The
reason is that Calabi-Yau condition fulfilled by ADE geometries is no longer
present for trivalent geometries. Restoring Calabi-Yau condition requires
promoting $T_{p,q,r}$ to a tetravalent geometry which we will denote as $%
T_{p,q,r,-s}$, see also figure 5. As this generalized geometry is behind
engineering of fundamental matter, let us say few more words on it.\bigskip

\begin{figure}[tbh]
\begin{center}
\epsfxsize=4cm \epsffile{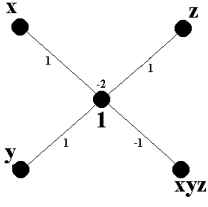}
\end{center}
\caption{{\protect\small \textit{This is a typical vertex for engineering \
fundamental matter in supersymmetric }}$\mathcal{N}=2${\protect\small
\textit{\ QFT}}$_{4}${\protect\small \textit{s. It may be noted as }}$%
T_{2,2,2,-2}$, \textit{the negative integer refers to promotion of original
trivalent geometry }$T_{2,2,2}$\textit{\ into a CY one.}}
\label{fig5}
\end{figure}
Generalized geometry $T_{p,q,r,-s}$ involves the following typical dimension
three vertices $V_{i}$,
\begin{equation}
V_{0}=\left( 0,0,0\right) ;\quad V_{1}=\left( 1,0,0\right) ;\quad
V_{2}=\left( 0,1,0\right) ;\quad V_{3}=\left( 0,0,1\right) ;\quad
V_{4}=\left( 1,1,1\right) ,  \label{4}
\end{equation}%
satisfying the following toric geometry relation
\begin{equation}
\sum_{i=0}^{4}q_{i}V_{i}=-2V_{0}+V_{1}+V_{2}+V_{3}-V_{4}=0  \label{5}
\end{equation}%
The vector charge $\left( q_{i}\right) =\left( -2,1,1,1,-1\right) $ is known
as the Mori vector and the sum of its $q_{i}$ components is zero as required
by the CY condition;
\begin{equation}
\sum_{i}q_{i}=0.
\end{equation}%
Before going ahead it is interesting to note that extension, of trivalent
geometry to Calabi Yau one, has an interpretation on the KM algebra side.
Let us illustrate this on analyzing corresponding "Cartan" matrices of
trivalent KM algebra and its CY extension. The promotion feature of
trivalent geometry $T_{2,2,2}$ into the CY $T_{2,2,2,-2}$ one is in fact a
very special extension. To $T_{2,2,2}$ vertex corresponds, on Lie algebraic
side, the following $4\times 4$ Cartan matrix%
\begin{equation}
K\left( T_{2,2,2}\right) =\left(
\begin{array}{cccc}
2 & 0 & -1 & 0 \\
0 & 2 & -1 & 0 \\
-1 & -1 & 2 & -1 \\
0 & 0 & -1 & 2%
\end{array}%
\right)
\end{equation}%
with $\det \left( T_{2,2,2}\right) =$ $4$; while for the CY extension, we
have the following $5\times 5$ matrix
\begin{equation}
K\left( T_{2,2,2,-2}\right) =\left(
\begin{array}{ccccc}
2 & 0 & -1 & 0 & 0 \\
0 & 2 & -1 & 0 & 0 \\
-1 & -1 & 2 & -1 & 1 \\
0 & 0 & -1 & 2 & 0 \\
0 & 0 & 1 & 0 & 2%
\end{array}%
\right) .  \label{ds}
\end{equation}%
Strictly speaking, $K\left( T_{2,2,2,-2}\right) $ is no longer a Cartan
matrix type since not all $K_{ij}$s, $i\neq j$, are negative integers as
required by eq(\ref{ne}). Following \cite{sa}, its determinant $\det \left(
T_{2,2,2,-2}\right) $ is given by%
\begin{equation}
\det \left( T_{2,2,2,-2}\right) =\det A_{1}\det T_{2,2,2}-\left( \det
A_{1}\right) ^{3},
\end{equation}%
which vanishes identically since $2\times 4-2^{3}=0$. Note also that $%
T_{2,2,2,-2}$ should be distinguished from $T_{2,2,2,2}$ which is a KM
algebra with a $5\times 5$ Cartan matrix given by,%
\begin{equation}
K\left( T_{2,2,2,2}\right) =\left(
\begin{array}{ccccc}
2 & 0 & -1 & 0 & 0 \\
0 & 2 & -1 & 0 & 0 \\
-1 & -1 & 2 & -1 & -1 \\
0 & 0 & -1 & 2 & 0 \\
0 & 0 & -1 & 0 & 2%
\end{array}%
\right)  \label{dd}
\end{equation}%
and whose determinant vanishes as well, $\det \left( T_{2,2,2,2}\right) =0$.
The difference between eqs(\ref{dd}) and (\ref{ds}) is that while in the
present case Dynkin vector is positive definite as shown below,%
\begin{equation}
\mathbf{d}=\left( 1,1,2,1,1\right) ,
\end{equation}%
its analog for $T_{2,2,2,-2}$ is no longer positive definite as is it given
by $\widetilde{\mathbf{d}}=\left( 1,1,2,1,-1\right) $. It follows then $%
T_{2,2,2,-2}$ does not fulfill Vinberg theorem requirements and so sits
beyond of KM classification.

\subsection{Mirror geometry}

In type IIB strings on mirror CY3, the $\left(
V_{0},V_{1},V_{2},V_{3},V_{4}\right) $ vertices are\ represented by complex
variables $\left( u_{0},u_{1},u_{2},u_{3},u_{4}\right) $ constrained as
\begin{equation}
\prod_{i}u_{i}^{q_{i}}=1
\end{equation}%
and solved by $\left( 1,x,y,z,xyz\right) $; see figure 5. In terms of these
variables, the algebraic eq describing mirror geometry, associated to eq(\ref%
{5}), is given by the following complex surface,
\begin{equation}
P\left( X^{\ast }\right) =\mathrm{e}+\mathrm{a}x+\mathrm{b}y+\left( \mathrm{c%
}-\mathrm{d}xy\right) z,  \label{6}
\end{equation}%
where $\mathrm{a,b,c,d}$ and $\mathrm{e}$ are complex moduli. Note that upon
eliminating the $z$ variable, the above (trivalent) algebraic geometry eq
reduces exactly to the standard bivalent vertex of $A_{1}$ geometry,
\begin{equation}
P\left( X^{\ast }\right) =\mathrm{a}x+\mathrm{e}+\frac{\mathrm{bc}}{\mathrm{d%
}}\frac{1}{x}.  \label{7}
\end{equation}%
The monomials $y_{0}=x$, $y_{1}=1$ and $y_{2}=\frac{1}{x}$ satisfy the well
known $su\left( 2\right) $ relation namely $y_{0}y_{2}=y_{1}^{2}$. To get
algebraic geometry eq of the CY3, one promotes the non zero moduli $\mathrm{%
a,b,c,d}$ and $\mathrm{e}$ to holomorphic polynomials on $\mathbb{CP}^{1}$
as follows:

\begin{eqnarray}
\mathrm{e}\left( w\right) &=&\sum_{i=0}^{n_{r}}\mathrm{e}_{i}w^{i};\qquad
\mathrm{a}\left( w\right) =\sum_{i=0}^{n_{r-1}}\mathrm{a}_{i}w^{i};\qquad
\mathrm{b}\left( w\right) =\sum_{i=0}^{n_{r+1}}b_{i}w^{i},  \notag \\
\mathrm{c}\left( w\right) &=&\sum_{i=0}^{m_{r}}c_{i}w^{i};\qquad \mathrm{d}%
\left( w\right) =\sum_{i=0}^{m_{r}^{\prime }}d_{i}w^{i};\qquad \mathrm{e}%
_{0},\mathrm{a}_{0},\mathrm{b}_{0},\mathrm{c}_{0},\mathrm{d}_{0}\neq 0.
\label{8}
\end{eqnarray}%
These analytic polynomials encode the fibrations of,%
\begin{equation}
SU\left( 1+n_{r-1}\right) \times SU\left( 1+n_{r}\right) \times SU\left(
1+n_{r+1}\right) ,
\end{equation}%
gauge invariance of the underlying $\mathcal{N}=2$ QFT$_{4}$ and its,
\begin{equation}
SU\left( 1+m_{r}\right) \times SU\left( 1+m_{r}^{\prime }\right) ,
\end{equation}%
flavor symmetry. These groups are engineered over the five nodes of the
extended trivalent vertex (CY 4-vertex). For instance $SU\left(
1+n_{r-1}\right) $ gauge symmetry is fibered over $V_{0}$ and $SU\left(
1+m_{r}\right) $ and $SU\left( 1+m_{r}^{\prime }\right) $\ flavor invariance
are fibered over the nodes $V_{3}$ and $V_{4}$ respectively; see figure
6.for illustration. Note in passing that $m_{r}^{\prime }$ refers to
engineering of fundamental matter on the extra node we have added to recover
CY condition.\bigskip

\begin{figure}[tbh]
\begin{center}
\epsfxsize=8cm \epsffile{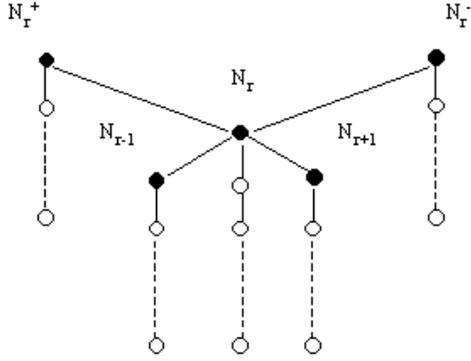}
\end{center}
\caption{{\protect\small \textit{This is a typical vertex of geometric
engineering of }}$\mathcal{N}\mathit{=2}$ {\protect\small \textit{%
supersymmetric QFT}}$_{4}$. $SU\left( 1+l_{i}\right) ${\protect\small
\textit{\ gauge and flavor symmetries are fibered over the five black nodes.
Flavor symmetries require large base volume.}}}
\label{fg6}
\end{figure}

\section{\textbf{Roots and 2-cycles in} $\mathcal{N}=2$ QFT$_{4}$s}

Here we consider the WL hyperbolic extension of affine ADE quiver gauge
theories. Since these hyperbolic algebras are simplest extensions of affine
ADE ones, one can easily make an idea on this larger structure just by
trying to extend known results.

\subsection{Simple roots in hyperbolic model}

We can write down several relations for hyperbolic ADE Lie algebras just by
using intuition and similarity arguments. Starting from a rank r affine
simply laced KM algebra with the usual simple roots $\alpha _{i},$ $%
i=0,1,...,r,$ and
\begin{equation}
\alpha _{0}=\delta -\sum_{i=1}^{r}d_{i}\alpha _{i}  \label{ab}
\end{equation}%
where $\delta $ stands for the usual imaginary root; $\delta ^{2}=0$, we can
immediately write down a realization of simple roots of the hyperbolic
extension. The rule is as follows: (\textbf{a}) Fix an affine ADE Cartan
matrix $K_{ij}^{0}=\alpha _{i}.\alpha _{j}$ and so a Dynkin diagram $%
\mathcal{D}^{0}$. We have for the case of affine A$_{r}$ the following
realization,
\begin{equation}
\alpha _{i}=e_{i}-e_{i+1};\qquad \alpha _{0}=\delta -\psi ,  \label{ac}
\end{equation}%
where $\left\{ e_{i}\right\} $ is a basis of orthonormal vectors. (\textbf{b}%
) Add an extra simple root $\alpha _{-1}$ to the game with the constraint
eqs,
\begin{equation}
\alpha _{-1}.\alpha _{-1}=2;\qquad \alpha _{-1}\alpha _{0}=-1;\qquad \alpha
_{-1}\alpha _{i}=0.  \label{ad}
\end{equation}%
(\textbf{c}) Because of the Lorentzian nature of the root lattice of
hyperbolic ADE algebras, this extra simple root is realized as,
\begin{equation}
\alpha _{-1}=\gamma -\delta ;\qquad \gamma ^{2}-2\gamma .\delta +\delta
^{2}=2,  \label{ae}
\end{equation}%
with $\gamma $ being a second basic imaginary root exhibiting the
properties:
\begin{eqnarray}
{\gamma }^{2} &=&{\delta }^{2}=0;\qquad {\gamma }{\delta }=-1,  \notag \\
{\gamma .}\alpha _{0} &=&1;\qquad {\gamma .}\alpha _{i}=0,\ i>0.  \label{af}
\end{eqnarray}%
Before proceeding, let us give some useful results: (\textbf{i}) Hyperbolic
ADE Lie algebras have two basic (light like) imaginary roots $\gamma $ and $%
\delta $ \ and so two special affine subalgebras. (\textbf{ii}) $\gamma $
and $\delta $ play a completely symmetric role. This symmetry is just a
specific Weyl transformation of root system in hyperbolic algebra. (\textbf{%
iii}) The symmetry between $\gamma $ and $\delta $ may be used to extract
part of information on hyperbolic structure that are not captured by affine
KM subsymmetry.

\subsection{2-Cycles in Calabi-Yau threefolds}

Root system $\left\{ \alpha \in \Delta \right\} $ of KM algebras have a
remarkable interpretation in algebraic geometry and in quiver gauge theories
embedded in type II string compactification on CY3 with ordinary and affine
ADE geometries. Following \cite{kmv1}, see also \cite{bs2}-\cite{bs4}, there
is a one to one correspondence between the following pairs: ADE algebra $%
\longleftrightarrow $ ADE geometry and ADE geometry $\longleftrightarrow $ $%
\mathcal{N}=2$ quiver gauge theories. More precisely, to each root $\alpha $%
, say the $su\left( n\right) $ simple root $\alpha _{i}=e_{i}-e_{i+1}$,
corresponds a two cycle $C$ (a basis 2-cycles $C_{i}$) of the ADE geometry
of CY3. We have the following 2-cycles intersection formula,%
\begin{equation}
C_{i}\cdot C_{i}=-K_{ij},
\end{equation}%
where $K_{ij}$ is exactly the Cartan matrix of ADE algebra. The holomorphic
volumes $v_{i}=t_{i}-t_{i+i}$ the 2-cycle $C_{i}$ basis are interpreted in
quiver gauge theories in terms of gauge coupling moduli as shown below,
\begin{equation}
g_{i}^{-2}\sim \left\vert t_{i}-t_{i+i}\right\vert .
\end{equation}%
Here $t_{i}$\ are complex moduli. In general, we have the following
correspondences: (\textbf{1}) ADE algebra $\longleftrightarrow $ ADE
geometry,
\begin{equation}
\begin{tabular}{|l|l|l|}
\hline
Algebra & $\qquad \longleftrightarrow \qquad $ & Geometry \\ \hline
\textit{Simple roots}$\text{\ }\alpha _{i}$ & $\qquad \longleftrightarrow
\qquad $ & $2$\textit{-Cycles}$\text{ }C_{i}=\mathbb{P}_{i}^{1}$ \\ \hline
\textit{Imaginary} $\delta $ & $\qquad \longleftrightarrow \qquad $ & $2$%
\textit{-Torus} $\mathbb{T}^{2}$ \\ \hline
$\text{\textit{Weyl transformations} }\omega $ & $\qquad \longleftrightarrow
\qquad $ & \textit{Picard group} \\ \hline
\end{tabular}%
,  \label{ade}
\end{equation}%
where $\omega $ belongs to $W_{affine}$; the affine Weyl group generated by
reflections $r_{\alpha }$ and translations $t_{\alpha }$. These
transformations, which generate two proper subgroups of $W_{affine}$, are
defined as,%
\begin{eqnarray}
\omega _{\alpha }\left( \beta \right) &=&\beta -2\frac{\left( \alpha ,\beta
\right) }{\alpha ^{2}}\alpha ,,  \notag \\
t_{\alpha }^{\delta }\left( \beta \right) &=&\beta +(\delta ,\beta )\alpha
-\left( \alpha ,\beta \right) \delta -\frac{\alpha ^{2}}{2}\left( \delta
,\beta \right) \delta ,
\end{eqnarray}%
and simplify for simply laced affine KM algebras. For $\beta \in \Delta
_{finite}$, the translation $t_{\alpha }^{\delta }\left( \beta \right) $
takes a simple form namely $\beta -\left( \alpha ,\beta \right) \delta $. (%
\textbf{2}) ADE geometry $\longleftrightarrow $ Quiver QFT$_{4}$

\begin{equation}
\begin{tabular}{|l|l|l|}
\hline
Geometry & $\qquad \longleftrightarrow \qquad $ & $\mathcal{N}=2$ Quiver
gauge theory \\ \hline
$Vol\left( C_{i}\right) $ & $\qquad \longleftrightarrow \qquad $ & $\text{%
\textit{Gauge} \textit{coupling} \textit{constants} }g_{i}^{-2}$ \\ \hline
$Vol\left( C_{0}\right) $ & $\qquad \longleftrightarrow \qquad $ & $%
g_{s}^{-1}=\exp \left( -\phi \right) $; \\ \hline
\textit{Weyl symmetry} & $\qquad \longleftrightarrow \qquad $ & $\text{%
\textit{Seiberg like duality}}$ \\ \hline
\end{tabular}%
\end{equation}%
where $\phi $ is dilaton. Following \cite{ms1,br}, the above correspondence,
which involves the triplet composed by \textit{roots} of affine KM algebra,
\textit{2-cycles} of ADE geometry and \textit{gauge moduli} in $4d$ $%
\mathcal{N}=2$ quiver gauge theories, may be extended naturally to
hyperbolic KM algebras and likely to the more general indefinite ones.
Similarly to above affine model, one also has a correspondence between
hyperbolic root system, 2-cycles and gauge moduli in hyperbolic gauge
theories. We have the following results:

\subsubsection{Hyperbolic algebra $\longleftrightarrow $ hyperbolic geometry}

To start recall that simple root basis of hyperbolic extension of affine ADE
algebras involves, in addition to $\left\{ \alpha _{0},\alpha _{i}\right\} $%
, an extra simple root $\alpha _{-1}$ whose Lie algebra features are given
by eqs(\ref{ae}-\ref{af}). The simple root system $\left\{ \alpha
_{-1},\alpha _{0},\alpha _{i}\right\} $ involves, amongst others, the two
imaginary roots $\gamma $ and $\delta $,%
\begin{equation}
\delta =\sum_{i=0}^{r}d_{i}\alpha _{i};\qquad \gamma
=\sum_{i=-1}^{r}d_{i}\alpha _{i},
\end{equation}%
where $d_{i}$, $0\leq i\leq r$\ are the usual Dynkin weights and where $%
d_{-1}$, which is equal to one, is their hyperbolic analog. Extending
results on type II string interpretation of affine models to the hyperbolic
ADE case, we see that geometrically speaking, $\gamma $ and $\delta $
generate two heterotic torii $T_{\pm }^{2}$,%
\begin{equation}
T_{+}^{2}=\sum_{i=0}^{r}d_{i}S_{i}^{2};\qquad
T_{-}^{2}=\sum_{i=-1}^{r}d_{i}S_{i}^{2},
\end{equation}%
where, roughly speaking, the $S_{i}^{2}$s are the usual real two spheres of
type II strings on CY3s with deformed ADE singularities. $S_{-1}^{2}$ is the
2-sphere associated with the root $\alpha _{-1}$.
\begin{equation}
\begin{tabular}{|l|l|l|}
\hline
Hyperbolic algebra & $\qquad \longleftrightarrow $ & hyperbolic geometry \\
\hline
\textit{Simple root }$\alpha _{-1}$ & $\qquad \longleftrightarrow $ &
\textit{2-sphere} $S_{-1}^{2}$ \\ \hline
\textit{Imaginary root }$\delta $ & $\qquad \longleftrightarrow $ & \textit{%
2-torus} $T_{+}^{2}$ \\ \hline
\textit{Imaginary root }$\gamma $ & $\qquad \longleftrightarrow $ & \textit{%
2-torus} $T_{-}^{2}$ \\ \hline
$\gamma ^{2}=\delta ^{2}=0,$ $\gamma \cdot \delta =-1$ & $\qquad
\longleftrightarrow $ & \textit{Intersection }$T_{\pm }^{2}\cdot T_{\pm
}^{2}=0,\quad $\textit{\ }$T_{+}^{2}\cdot T_{-}^{2}=-1$ \\ \hline
\end{tabular}
\label{h}
\end{equation}%
Note that $T_{+}^{2}$ and $T_{-}^{2}$ may be thought of as holomorphic $%
T_{z}^{2}$ and anti-holomorphic $T_{\overline{z}}^{2}$ two cycles in Kahler
threefolds. This observation constitutes a key point in our quest for
searching indefinite symmetries in hyperbolic ADE quiver gauge models.

\subsubsection{Hyperbolic geometry $\longleftrightarrow $ hyperbolic gauge
model}

Extending results on type IIB superstring interpretation of 4d $\mathcal{N}%
=2 $ affine quiver gauge theories, it is not difficult to see that the
following correspondence holds,
\begin{equation}
\begin{tabular}{|l|l|l|}
\hline
hyperbolic geometry & $\qquad \longleftrightarrow \qquad $ & hyperbolic $%
\mathcal{N}=2$ QFT$_{4}$ \\ \hline
$C_{i},\quad i=1,...$ & $\qquad \longleftrightarrow \qquad $ & $\text{%
\textit{Gauge} \textit{coupling} \textit{constants} }g_{i}^{-2}$ \\ \hline
$T_{+}^{2}=\sum_{i\geq 0}d_{i}C_{i}$ & $\qquad \longleftrightarrow \qquad $
& $g_{+}^{-1}=exp(-\phi )+\chi $ \\ \hline
$T_{-}^{2}=\sum_{i\geq -1}d_{i}C_{i}$ & $\qquad \longleftrightarrow \qquad $
& $g_{-}^{-1}=exp(-\phi )-\chi $ \\ \hline
\textit{Weyl symmetry} & $\qquad \longleftrightarrow \qquad $ & \textit{%
generalized} $\text{\textit{Seiberg like duality}}$ \\ \hline
\end{tabular}%
\end{equation}%
Like for $Vol\left( C_{0}\right) $, this correspondence allows us to
interpret $Vol\left( C_{-1}\right) $ in terms of type IIB string moduli
space, except now that we should have\ a non zero action $\chi $. Note that,
upon performing a Wick rotation of the RR axion moduli ($\chi
\longrightarrow i\chi $), the inverse string coupling $g_{+}^{-1}$ becomes $%
g_{\tau }^{-1}=exp(-\phi )-i\chi $ which should be compared with the complex
moduli $i\tau _{IIB}$ of the holomorphic two torus used in F-theory
compactification on elliptic K3 \cite{fth}.

\section{Hyperbolic invariance: Two examples}

In this section, we present two field theoretical models where hyperbolic
invariance appears as a symmetry. The first example deals with the complete
classification of $\mathcal{N}=2$\textbf{\ }CFT$_{4}$ in four dimensions.
The second example concerns $\mathcal{N}=2$ hyperbolic quiver gauge theories
embedded in type IIB superstrings on CY3s; but with non zero axion.

\subsection{Example I: Classification of $N=2$\ CFT$_{4}$s}

Four dimensional $N=2$ supersymmetric CFTs constitute an important class of
QFT$_{4}$s embedded in type II superstring compactifications on elliptic
fibered CY threefolds with $ADE$ geometries preserving eight
supersymmetries. These special field theoretic models give exact solutions
for the moduli space of the Coulomb branch of $\mathcal{N}=2$ QFT$_{4}$s and
admit a very nice geometric engineering in terms of quiver diagrams. Common $%
\mathcal{N}=2$ CFT$_{4}$s were first believed to be classified into two
categories according to the type of singularities; but as we will see, there
are in fact three categories classified by Vinberg theorem.

\textbf{(1)} Ordinary super CFT$_{4}$s:$\qquad \mathcal{N}=2$ supersymmetric
CFT$_{4}$ based on \textit{finite} $ADE$ singularities with quiver gauge
group,
\begin{equation}
G=\prod_{i}SU\left( n_{i}\right) \times G_{flavor}
\end{equation}%
and matter in both fundamental $\mathbf{n}_{i}$ and bi-fundamental $\left(
\mathbf{n}_{i}\mathbf{,}\overline{\mathbf{n}}_{j}\right) $ representations
of $G$.

\textbf{(2)} Affine super CFT$_{4}$s:$\qquad \mathcal{N}=2$ CFT$_{4}$ with
quiver gauge group
\begin{equation}
G=\prod_{i}SU\left( d_{i}n\right) ,
\end{equation}%
and bi-fundamental matter only. This second category of scale invariant
field models are classified by \textit{affine} $ADE$ algebras. The positive
integers $d_{i}$ appearing in $G$ are the usual Dynkin weights considered
earlier; they form a special positive definite integer vector $\mathbf{d}%
=\left( d_{i}\right) $ satisfying $\mathbf{K}_{ij}^{0}d_{j}=0$ and so
\begin{equation}
\mathbf{K}_{ij}^{0}n_{j}=0.  \label{1}
\end{equation}%
Here $n_{j}=nd_{j}$ and $\mathbf{K}^{0}$ is the affine Cartan matrix. Before
describing the third class, let us comment a little bit the emergence of eq(%
\ref{1}) in the CFT game.\ Though not surprising, the appearance of this
remarkable eq in the geometric engineering of $\mathcal{N}=2$ CFT$_{4}$s is
very exciting. \textit{First} because $4d$ conformal invariance requiring
the vanishing of the total holomorphic beta function $b,$ which requires in
turn the vanishing of the beta function factors $b_{i}$,%
\begin{equation}
b_{i}=\frac{1}{12}\left( 44n_{i}-2\sum_{j}\left[ 4a_{ij}^{4}+a_{ij}^{6}%
\right] n_{j}\right) ,\qquad 3a_{ij}^{4}=3\overline{a}_{ij}^{4}=2a_{ij}^{6},
\end{equation}%
is now translated into a condition on allowed KM algebras eq(\ref{1}). In
above relation $a_{ij}^{4}$ is the number of fundamental fermions and $%
a_{ij}^{6}$ the number of adjoint scalars. \textit{Second}, even for $%
\mathcal{N}=2$ CFT$_{4}$ based on \textit{finite} $ADE$ with $m_{i}\mathbf{n}%
_{i}$ fundamental matters, the condition for scale invariance may be also
formulated in terms of the corresponding Cartan matrix $\mathbf{K}^{+}$\ as,
\begin{equation}
\mathbf{K}_{ij}^{+}n_{j}=m_{i}.  \label{2}
\end{equation}%
Note that identities (\ref{1},\ref{2}) can be rigorously derived by starting
from mirror geometry of type IIA string on Calabi-Yau threefolds and taking
the field theory limit in the weak gauge couplings $g_{r}$ regime associated
with large volume of the CY3 base ( $V_{r}=1/\varepsilon $ with $\varepsilon
\rightarrow 0$ ). In this limit and using eqs(\ref{6}-\ref{8}) and figure 6,
one shows that complex deformations $a_{r,l}$, $b_{r,l}$, $c_{r,l}$, $%
d_{r,l} $ and $e_{r,l}$ of the mirror geometry scale as,
\begin{eqnarray}
a_{r,l} &=&\varepsilon ^{l-n_{r-1}},\qquad b_{r,l}=\varepsilon
^{l-n_{r+1}},\qquad c_{r,l}=\varepsilon ^{l-m_{r}},  \notag \\
d_{r,l} &=&\varepsilon ^{l-m_{r}^{\prime }},\qquad e_{r,l}=\varepsilon
^{l-n_{r}},  \label{dm}
\end{eqnarray}%
where $m_{r}^{\prime }$ is taken non zero for later consideration. One shows
moreover that the universal coupling parameters $Z\left( g_{r}\right) $
given by,%
\begin{equation}
Z\left( g_{r}\right) =\frac{a_{r,0}b_{r,0}c_{r,0}}{e_{r,0}^{2}d_{r,0}},
\end{equation}%
behave as $\varepsilon ^{-b_{r}}$ with $b_{r}$s, the beta function
components of the $r$-th $U\left( n_{r}\right) $\ gauge subgroup factor of $%
G $; given by,%
\begin{equation}
b_{r}=2n_{r}-n_{r-1}-n_{r+1}-\left( m_{r}-m_{r}^{\prime }\right) .
\label{br}
\end{equation}%
In the limit $\varepsilon \rightarrow 0$, finiteness of universal couplings $%
Z^{\left( g_{r}\right) }$ requires $b_{r}\leq 0$; i.e the field theory limit
should be asymptotically free. Scale invariance of the $\mathcal{N}=2$ CFT$%
_{4}$s requires however $b_{r}=0$ \ for all $r$ indices. Upon setting%
\begin{equation}
u_{i}=n_{i};\qquad v_{i}=\left\vert m_{i}-m_{i}^{\prime }\right\vert ,
\label{dn}
\end{equation}%
it is not difficult to see that eq(\ref{br}) coincide exactly with eqs(\ref%
{1},\ref{2}). The vanishing condition of $b_{r}$s is then translated into a
condition on the intersection matrix ( Cartan matrix) $K_{ij}^{\left(
q\right) }$ of $ADE$ singularities one is considering; i.e finite, affine or
indefinite. As a result, one sees that, along with ordinary and affine
supersymmetric CFT$_{4}$s, there is moreover an extra remarkable indefinite
sector. In what follows, we will give comments on indefinite CFTs; for more
details and technicalities see \cite{kmv1,ms1,ms2}.

\textbf{(3)} Indefinite $\mathcal{N}=2$ CFT$_{4}$s:$\qquad $ To start, note
that appearance of eqs(\ref{1},\ref{2}) can be viewed as just the two
leading relations of a more general result associated with the triplet,%
\begin{equation}
\mathbf{K}_{ij}^{q}n_{j}=qv_{i};\qquad q=+1,0,-1.  \label{3}
\end{equation}%
From this view, one already suspects that $\mathcal{N}=2$ CFT$_{4}$s should
be classified by Vinberg formula eq(\ref{d}). At a first sight, this
relation tells us, amongst others, that: (\textbf{i}) there are three main
classes of $\mathcal{N}=2$ CFT$_{4}$s, (\textbf{ii}) the extra indefinite $%
\mathcal{N}=2$ CFT$_{4}$ sector is given by,%
\begin{equation}
\mathbf{K}_{ij}^{-}n_{j}=-v_{i}.
\end{equation}%
Following \cite{ms1,ms2,br} and using above analysis eqs(\ref{dm}-\ref{dn}),
one shows that eq(\ref{3}) do follow indeed from type II string on
Calabi-Yau threefolds with geometries having a 2-cycle basis $\left\{
C_{i}\right\} $ satisfying,%
\begin{equation}
C_{i}\cdot C_{i}=\mathbf{K}_{ij}^{q};\qquad q=+1,0,-1.  \label{q}
\end{equation}%
The derivation of eq(\ref{3}) is then proved; it relies on using trivalent
geometry (figure 6) and proceeds as in eq(\ref{6}-\ref{8}) by considering
the \textit{general representation} of fundamental engineering fundamental
matter. Let re-comment briefly the main steps by using different, but
equivalent, sentences: (\textbf{i}) Use geometric engineering of $\mathcal{N}%
=2$ QFT$_{4}$s described in section 3. (\textbf{ii}) Fundamental matter
require extension of trivalent vertices with the five entry Mori vectors,
\begin{equation}
q_{\mathrm{\tau }}=\left( 1,-2,1;1,-1\right) .
\end{equation}%
The first three ones namely $\left( 1,-2,1\right) $ are common entries since
they are involved in the geometric engineering of gauge fields and
bi-fundamental matter. They lead to $\mathbf{K}_{ij}^{0}n_{j}=0$. The fourth
entry is used in the engineering of fundamental matter of CFT$_{4}$s based
on \textit{finite} $ADE$ and lead to $\mathbf{K}_{ij}^{+}n_{j}=m_{i}$. The
fifth entry has been treated for sometimes as a spectator only needed to
ensure the CY condition $\sum_{\tau =1}^{5}q_{\mathrm{\tau }}=0$. Handling
this vertex on equal footing as the four previous others; i.e $m\prime
_{i}\neq 0$ in eq(\ref{dm}), gives surprisingly the missing third sector of
eqs(\ref{3}).

\textbf{Result:\qquad }$\mathcal{N}=2$ CFT$_{4}$s, embedded in type II
string compactification on Calabi-Yau threefolds with generalized ADE
geometries satisfying eqs(\ref{q}), are classified by Vinberg classification
theorem of KM algebras.

\subsection{Example II: Axion as a hyperbolic moduli}

To begin recall that 4d supersymmetric ADE quiver gauge theories are QFT$%
_{4} $ limits of type II strings on CY threefolds with ADE geometries and
are remarkably engineered on ADE Dynkin diagrams. Nodes of the Dynkin graphs
encode gauge and adjoint matter multiplets. Links between the nodes engineer
bi-fundamental matter involved in supersymmetric $\prod_{i}U\left(
N_{i}\right) $ quiver gauge theory. Recall also that roots $\alpha $ of KM
algebras,
\begin{equation}
\alpha =\pm \sum_{i}k_{i}\alpha _{i},\qquad k_{i}\in \mathbb{Z}_{+},
\end{equation}%
which are generated by the $\alpha _{i}$ simple ones, are generally realized
in $\mathbb{R}^{n}=\sum \mathbb{R}e_{i}$ in terms of specific linear
combinations of the $\left\{ e_{i}\right\} $ basis, $e_{i}e_{j}=\delta _{ij}$%
. For the case of $su\left( n\right) $ simple roots for instance, we have
\begin{equation}
\alpha _{i}=e_{i}-e_{i+1},\qquad i=1,....  \label{ee}
\end{equation}%
Together with this eq, there are also relations such that $\alpha
_{1}+...+\alpha _{n}=e_{1}-e_{n+1}$ which are useful in the study of quiver
gauge theories. With these results in mind; one can go ahead to study
supersymmetric QFT$_{4}$s embedded in type II strings on CY3 with ADE
geometries and their hyperbolic extension. \ To that purpose, recall that
roots have algebraic geometry analogs in CY3 with generalized ADE
geometries. Some of the results regarding this correspondence were already
shown on eqs(\ref{ade}-\ref{h}). In what follows, we complete the picture of
subsection 5.1 by giving details and comments.

One of the relevant objects for our study concerns the correspondence
between roots of simply laced KM algebra and 2-cycles of the associated
geometry. Of particular interest are simple roots $\alpha _{i}$ which are in
one to one correspondence with the holomorphic "volumes" $\zeta _{i},$
\begin{equation}
\zeta _{i}=\int_{\mathbb{P}_{i}^{1}}\Omega ^{\left( 2,0\right) }.
\end{equation}%
In this eq, $\Omega ^{\left( 2,0\right) }$ is the usual holomorphic two form
on the complex dimension one projective space $\mathbb{P}^{1}$. So the $%
\zeta _{i}$s are the volumes of the homological two-cycles $\mathbb{P}%
_{i}^{1}\sim S_{i}^{2}$ involved in the deformation of ADE singularities.
Let us make two comments regarding these complex holomorphic volumes $\zeta
_{i}$. First, note that their explicit values are given by,
\begin{equation}
\zeta _{i}=t_{i}-t_{i+1},\qquad i=1,....,
\end{equation}%
where $t_{i}$s are complex moduli. These values should be compared with eq(%
\ref{ee}) and allow to write down holomorphic volumes of any 2-cycles of the
ADE geometry by just using the analogy of roots. Second the $\zeta _{i}$s,
which describe deformations of local ADE geometry, have a nice
interpretation in $4d$ $\mathcal{N}=2$ supersymmetric quiver $\Pi
_{i=1}^{r}U\left( N_{i}\right) $ gauge theories with adjoint matter
superfields $\left\{ \Phi _{i}\quad i=1,...,r\right\} $. They appear as FI
like coupling constants generating the following $4d$ $\mathcal{N}=1$ linear
chiral superspace potential deformation $\mathrm{\delta W}$,
\begin{equation}
\mathrm{\delta W}=\sum_{i=1}^{r}\zeta _{i}\int d^{4}xd^{2}\theta Tr\left(
\Phi _{i}\right) \text{.}  \label{dw}
\end{equation}%
This special superpotential deformation preserves $\mathcal{N}=2$
supersymmetry and its non linear $\mathcal{N}=1$ extension $\mathrm{\delta W}%
\sim \Phi ^{n+1}$ is at the basis of the field theoretic representation of
the geometric transition $O\left( -1\right) \times O\left( -1\right) \times
\mathbb{CP}^{1}\rightarrow T^{\ast }S^{3}$ of the conifold. Eq(\ref{dw}) is
also behind the field theoretic analysis of large $N$\ field dualities and
in the derivation of exact results in $\mathcal{N}=1$ supersymmetric gauge
theories.

The other relevant object we want to give here concerns Weyl group rotating
roots of hyperbolic KM algebra and duality symmetry in hyperbolic quiver
gauge theories. For affine ADE KM algebras, it is now well established that
Weyl group $W_{ADE}$ is associated with Seiberg like dualities and its
translation subgroup is behind the RG cascades of affine models. At low
energies below string scale where the dynamics of matter and gauge fields is
governed by supersymmetric Yang Mills model, one disposes of sets\ of dual
ADE quiver gauge theories with a remarkable subclass whose duality
symmetries act on previous $\upsilon _{i}$ as
\begin{equation}
\upsilon _{i}\rightarrow \upsilon _{i}^{\prime }=A_{ij}\upsilon _{j},
\label{v}
\end{equation}%
These duality symmetries were shown to be isomorphic to the usual Weyl group
transformations of ADE root system \cite{c8}. By help of correspondence (\ref%
{v}), the $A_{ij}$\ matrix in above relation is isomorphic to the bi-linear
product\ $\delta _{ij}-\alpha _{i}\alpha _{j}$ that appears in Weyl
reflections,
\begin{equation}
\alpha _{i}^{\prime }=\alpha _{i}-2\frac{\alpha _{i}\alpha _{j}}{\alpha
_{j}^{2}}\alpha _{j}.
\end{equation}%
In addition to the two above links, there are other basic links between $4d$
super quiver QFT$_{4}$s and ADE algebra. For instance ADE root systems and
their Weyl symmetries are also used in brane realization of the quiver gauge
theories living in the world volume of parallel $N_{0}$ D3 branes and $N$ D5
ones partially wrapping CP$_{i}^{1}$ two-cycles of CY3 folds with a local
ADE geometry. There, $N_{0}$ D3 are roughly speaking associated with the
affine simple root $\alpha _{0}$ of affine KM ADE root system and wrapped $N$
D5s with remaining $\alpha _{i}$ simple ones. In this representation, field
theoretic scenarios such as higgsings correspond just to special properties
of the root system. Other basic relations between roots and their Weyl
automorphisms on one hand; and relevant QFT$_{4}$ moduli on the other hand
can be also written down. Supersymmetric Yang-Mills gauge couplings g$%
_{i}^{SYM}$ of the quiver gauge subgroup factors $U\left( N_{i}\right) $ and
corresponding beta functions $b_{i}$ including supersymmetric affine ADE
conformal field models,
\begin{equation}
\frac{1}{g_{s}}=\sum_{i=0}^{r}d_{i}g_{i}^{-2};\qquad
b_{D}=\sum_{i=0}^{r}d_{i}b_{i},
\end{equation}%
obey a similar law as holomorphic volumes $\upsilon _{i}$ eq(\ref{v}). For
details on this issue as well as other areas of involvement of Weyl
symmetries, we refer to \cite{br,d}, see also \cite{d0,d1,d2,d3}. To get the
relation between $g_{\pm }$ and the $g_{i}$ moduli of the hyperbolic quiver
gauge theory, it is enough to recall the relation between the string
coupling $g_{s}$ and the $g_{i}$ gauge couplings in the affine model.
\begin{equation}
\mathrm{g}_{s}^{-1}|_{\chi =0}=g_{s}^{-1}=\sum_{i=0}^{r}d_{i}g_{i}^{-2},
\label{gs}
\end{equation}%
For non zero $\chi $, instead of eq(\cite{gs}) we have rather,
\begin{equation}
g_{-}^{-1}=\sum_{i=0}^{r}d_{i}g_{i}^{-2};\qquad
g_{+}^{-1}=\sum_{i=-1}^{r}d_{i}g_{i}^{-2};
\end{equation}%
where $g_{-1}$ is the gauge coupling of the gauge group engineered on the
hyperbolic node of the Dynkin diagram $\mathcal{D}^{-}$. From above
relation, we also see that the parameters $g_{\pm }$ are linked as follows,
\begin{equation}
g_{+}^{-1}=g_{-1}^{-2}+g_{-}^{-1},
\end{equation}%
Substituting $g_{+}^{-1}$ and $g_{-}^{-1}$ in terms of the dilaton $\phi $
and the axion $\chi $ expressions, we get
\begin{equation}
g_{-1}^{-2}=2\chi \qquad \Longrightarrow \qquad \chi \sim vol\left(
C_{-1}\right)
\end{equation}%
Taking $\chi =0$, one recovers the usual affine model with all desired
features. The main result of this subsection is that the volume $\upsilon
_{-1}$ of hyperbolic 2-cycle $C_{-1}$ is given by axion modulus $\chi $. In
the limit $\chi \rightarrow 0$, the corresponding gauge group factor becomes
a flavor symmetry.

\section{Conclusion and discussion}

In this paper, we have reviewed aspects on the classification of KM algebras
and presented two examples of field theoretical models exhibiting indefinite
KM symmetries. These field theories are obtained as limits of type II
superstring compactification on Calabi-Yau threefolds with generalized ADE
geometries. They concern the two following: \newline
(\textbf{1}) the classification of $\mathcal{N}=2$ CFT$_{4}$s into three
main subsets following from solving scale invariance of the more general $%
\mathcal{N}=2$ QFT$_{4}$ geometric engineering condition namely,%
\begin{equation}
b_{i}=K_{ij}^{q}n_{j}-q\left\vert m_{i}-m_{i}^{\prime }\right\vert =0,
\label{bb}
\end{equation}%
where $m_{i}$ and $m_{i}^{\prime }$ are numbers of fundamental matter. With
this result, the general picture on full classification of $\mathcal{N}=2$
CFT$_{4}$s is as follows: (\textbf{i}) Ordinary $\mathcal{N}=2$ CFT$_{4}$s
classified by finite ADE Lie algebra, (\textbf{ii}) Affine $\mathcal{N}=2$
CFT$_{4}$s associated with affine KM symmetries and (\textbf{iii})
Indefinite $\mathcal{N}=2$ CFT$_{4}$s described by indefinite KM algebras.
In present study, we have considered only simply laced KM algebras; it would
be interesting to check whether this result is valid as well for non simply
laced KM models.

Before proceeding, we would like to add two more things which have not been
discussed in the paper. First, it should be noted that as far as type II
superstring compactification on Calabi Yau threefolds is concerned, one
learns from present study that there three classes of CY3 with K3
fibrations. These threefolds, which are also classified by Vinberg theorem,
have 2-cycle basis $\left\{ C_{i}\right\} $ with an intersection matrix
given by,%
\begin{equation}
C_{i}\cdot C_{i}=K_{ij}^{q};\qquad q=1,0,-1,
\end{equation}%
where $K_{ij}^{q}$ is as specified in the development of the present paper.
Second, it is interesting to note that as far as affine models are
concerned, one should distinguish between to cases. ($\alpha $) Usual affine
model involving Dynkin diagram of affine KM algebras; they correspond to set
$m_{i}=m_{i}^{\prime }=0$ in eq(\ref{bb}). ($\beta $) Affine model involving
"trivalent" geometries with the condition $m_{i}=m_{i}^{\prime }\neq 0$.%
\newline
(\textbf{2}) hyperbolic quiver gauge theories embedded in type IIB string on
specific CY3s. Considering CY threefolds having 2-cycle $\left\{
C_{i}\right\} $ basis with the following intersection formula,%
\begin{equation}
C_{i}\cdot C_{j}=K_{ij}^{-},\qquad i,j=-1,0,2,...,
\end{equation}%
where $K^{-}$ is a generic hyperbolic Cartan matrix, we have shown that
results on affine models extend naturally to the hyperbolic case. In
particular, we have shown that the volume of $C_{-1}$, the cycle associated
with the hyperbolic simple root, is given by axion $\chi $. This means that
for non zero axion, there is a $U\left( N_{-1}\right) $ gauge symmetry with
non zero gauge coupling $g$ engineered on the hyperbolic node of the
underlying CY geometry. More details on this issue as well as other results
on hyperbolic quiver QFT$_{4}$s such as RG cascades may be found in \cite{br}
and refs therein.

In the end of this discussion, we should note that this study may be viewed
as a first step towards the exploration of indefinite KM algebras. Besides
field theory limits, it also opens an issue for looking for hidden
indefinite symmetries in type II superstring theory and on the
classification of generalized ADE geometries in CY3 \cite{aa}.

\begin{acknowledgement}
I would like to thank the organizers of IPM String School and Workshop
ISS2005, Qeshm Island, Iran,\ for generosity kind hospitality. Special thank
to\ M. Alishahiha, F Ardalan and S. Sheikh-Sebbari for fruitful discussions
and IPM staff for help. I would like to thank also R. Ahl Laamara, M. Ait
Ben Haddou, A Belhaj and B.L. Drissi for earlier collaboration on this
matter. This research work is undertaken in the framework of Protars III
D12/25 CNRST.
\end{acknowledgement}

\end{document}